\documentclass[a4paper,fleqn,usenatbib]{mnras}
\usepackage[T1]{fontenc}
\usepackage{ae,aecompl}

\usepackage{graphicx}	% Including figure files
\usepackage{amsmath}	% Advanced maths commands
\usepackage{amssymb}	% Extra maths symbols
\usepackage{times}   
\usepackage{natbib}   
\usepackage{color,ulem}  
\bibpunct{(}{)}{;}{a}{}{,}

\title[NGC\,3610 and its globular cluster system]{The merger remnant NGC\,3610 and its 
globular cluster system: \\ a large-scale study}

\author[L. P. Bassino et al.]{ 
Lilia P. Bassino$^{1,2}$\thanks{E-mails: lbassino@fcaglp.unlp.edu.ar (LPB); 
jpcaso@fcaglp.unlp.edu.ar (JPC)} and Juan P. Caso$^{1,2}$
\\
$^{1}$Facultad de Ciencias Astron\'omicas y Geof\'isicas de la Universidad Nacional de La Plata, and  
Instituto de Astrof\'isica de La Plata \\ (CCT La Plata -- CONICET, UNLP), Paseo del Bosque S/N, 
B1900FWA La Plata, Argentina\\   
$^{2}$Consejo Nacional de Investigaciones Cient\'ificas y T\'ecnicas, Rivadavia 1917, C1033AAJ  
Ciudad Aut\'onoma de Buenos Aires, Argentina}

\date{Accepted XXX. Received YYY; in original form ZZZ}

\pubyear{2016}

\begin{document}
\label{firstpage}
\pagerange{\pageref{firstpage}--\pageref{lastpage}}
\maketitle

\begin{abstract}

We present a photometric study of the prototype merger remnant NGC\,3610 and its globular cluster (GC) 
system, based on new GEMINI/GMOS and ACS/HST archival images. Thanks to the large FOV of 
our GMOS data, larger than previous studies, we are able to detect a `classical' bimodal GC colour 
distribution, corresponding to metal-poor and metal-rich GCs, at intermediate radii and a small subsample 
of likely young clusters of intermediate colours, mainly located in the outskirts. The extent of the 
whole GC system is settled as about 40\,kpc.  
The GC population is quite poor, about $500\pm110$\,members  
that corresponds to a low total specific frequency $S_N \sim 0.8$. The effective radii of a cluster sample 
are determined, including those of two spectroscopically confirmed young and metal-rich 
clusters, that are in the limit between GC and UCD sizes and brightness. The large-scale galaxy 
surface-brightness profile can be decomposed as an inner embedded disc and an outer spheroid, 
determining for both larger extents than earlier research (10\,kpc and 30\,kpc, respectively). 
We detect boxy isophotes, expected in merger remnants, and show a wealth of fine-structure in the 
surface-brightness distribution with unprecedented detail, coincident with the outer spheroid. 
The lack of symmetry in the galaxy colour map adds a new piece of evidence to the recent merger 
scenario of NGC\,3610.  
\end{abstract}


\section{Introduction} 

Globular cluster systems (GCSs) have been extensively studied in the 
past two decades, but a great effort has been focused mainly in early-type 
galaxies in dense environments. The large number of globular clusters (GCs) 
present in these systems 
\citep[e.g.][and references therein]{dir03a,bas06a,har13} favours the 
statistical study of GCs, but restricts our knowledge to these particularly
evolved systems. Despite early-type galaxies in low environments usually have
rather poor GCSs \citep[e.g.][]{spi08,lan13,cas13b}, their characteristics
might provide clues to figure out the evolutionary history of the galaxy
itself \citep[e.g.][]{sal15,esc15,cas15a}.

NGC\,3610 is considered as a prototype of an intermediate-age merger remnant \citep{how04}, 
classified as peculiar lenticular or shell elliptical galaxy \citep[e.g.][and NED, NASA/IPAC 
Extragalactic Database]{dev91}. The presence of a `fine-structure' of 
shells, plumes, boxy isophotes, and `X-features' is widely considered as typical of disc-disc 
mergers of similar mass galaxies \citep[e.g.][]{her92,bar92}. Located in a low-density 
environment, NGC\,3610 is included in the group LGG\,234
\citep{gar93} or the NGC\,3642 group \citep{fou92}, in both cases
a five-member group composed of the same galaxies. In fact, \citet{mad04} 
identified four physical companions (with known redshifts) within 300\,kpc 
and $\pm 225$\,km\,sec$^{-1}$ from the central galaxy.  

The complex structure of NGC\,3610 has been studied for decades. 
In their analysis of about 160 early-type galaxies, \citet{ebn88} had 
already detected that NGC\,3610 posses boxy isophotes. 
The structure of NGC\,3610 has originally been studied by \citet{sco90}, 
who pointed out at the unusual presence of an inner disc 
in an elliptical galaxy, while \citet{sch92} selected this galaxy 
as the one presenting the richest fine-structure among their sample 
of 69 E and S0 merger candidates. \citet{mic94} studied the morphology 
of over 100 E-S0 galaxies and pointed that NGC\,3610 has a strongly 
twisted and peculiar envelope as well as an asymmetric structure, 
presenting an embedded inner disc. Later on, \citet{whi97} performed 
a complementary study of the galaxy and its GCS based on HST data, 
presenting evidence that it is a dynamically young elliptical. 
 
\citet{whi97} also pointed to the existence of an intermediate age 
population of GCs in NGC\,3610, that might have originated in a past event related to 
the peculiar structure of this galaxy. Afterwards, the inner GCS was studied by  
\citet{whi02} and \citet{gou07} on the basis of images obtained with the Wide Field 
Planetary Camera 2 and the Advanced Camera for Surveys (ACS), respectively, both on board 
the Hubble Space Telescope (HST). They revealed an unusual behaviour of the red metal-rich  
GC luminosity function (GCLF), being this sub-population of intermediate age ($\sim 1.5 -4$\,Gyr)  
and formed during a gas-rich merger. The inner half of such sub-population (for radii 
smaller than $\sim 45$\,arcsec) showed a flattening in the GCLF that is consistent with the 
predictions of GC disruption models \citep[e.g.][]{fal01}. Thus, it is a likely consequence of the 
stronger tidal field in the inner regions, that cause a more effective low-mass cluster disruption 
than further out. 
Alternatively, the outer half (up to a galactocentric distance $\sim 100$\,arcsec) 
is consistent with a power-law GCLF \citep[see also][]{gou04}.

In addition, a small sample of GCs has been spectroscopically confirmed 
\citep{str03,str04b}, including a couple of young and metal-rich ones. They are 
identified as W6 and W11, with ages of 1--2\,Gyr and 1--3\,Gyr old, and  
metallicities $[Z/H] =$ +0.4 and +0.7, respectively. These ages are in agreement  
with the estimate of 1.6 +/- 0.5\,Gyr obtained for NGC\,3610 by \citet{den05}. 
According to such result, it is inferred that   
the intermediate-age clusters formed in the disc-disc merger that has probably originated 
the galaxy. Later on, \citet{geo12} obtained photometrical ages and metallicities for 
a sample of 50 bright GC candidates (some of them already studied by Strader et al.), 
using optical and near-IR imaging. By means of colour-colour diagrams based on 
that photometric combination, it is possible to break the age-metallicity degeneracy. 
Comparing with the spectroscopically derived parameters, the metallicities are in agreement 
while photometric ages are $\sim 2$\,Gyr older than spectroscopic ones, though the age 
difference becomes smaller for more metal-rich GCs. The age and metallicity distributions 
obtained by \citet{geo12} also point to their bright cluster sample being dominated by a 
metal-rich and intermediate-age sub-population. 

This paper has been planned as a complement to the previous ones, as our observational 
data will allow us to cover one of the largest field-of-view (FOV) used to study this 
galaxy and its GCS so far. Thus, we will be able 
to establish the whole GCS extension, perform an homogeneous comparison of the GC colour 
distributions in different radial regimes, as well as analyse the 
outer regions of the surface-brightness distribution of the galaxy itself.

The most recent distance determinations for NGC\,3610, based on surface-brightness fluctuations 
(SBF) gives $\sim 35$\,Mpc \citep{can07,tul13}, remarkably larger than previous ones 
\citep{ton01,bla01}. Its heliocentric radial velocity is $1707\pm5$km\,s$^{-1}$ \citep{cap11}, 
lower than expected for the redshift independent distances. In the following, we will adopt the  
recent SBF estimates, hence a distance modulus $m-M = 32.7\pm0.1$.

This paper is organized as follows. The observations and data reduction are presented 
in section\,2, the results are described in section\,3, and section\,4 is 
devoted to the discussion. Finally, a summary and the concluding remarks are given 
in section\,5.

\section{Observations and reduction}

\subsection{Observational data}
The data set consists of images obtained with GMOS (Gemini North) 
in $g'$, $r'$, and $i'$ filters, during semester  
2013A (programme GN2013A-Q-42, PI: J.P. Caso), plus ACS (HST) 
images in $F555W$ and $F814W$ filters (programme 9409, PI: P. Goudfrooij) obtained
from the HST Archive, and originally observed during June 2003. 

\begin{figure}    
\includegraphics[width=85mm]{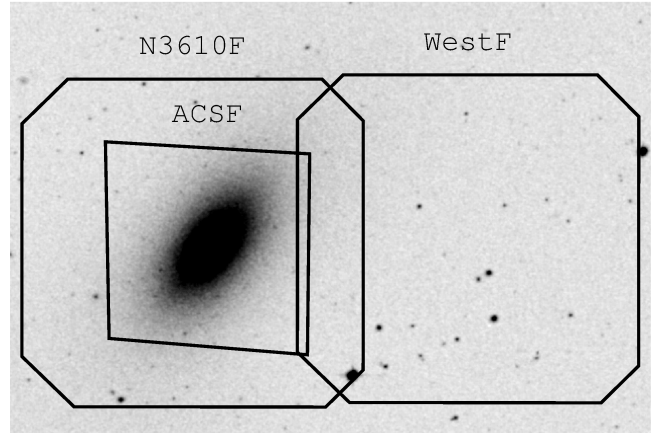}    
\caption{Two GMOS-S fields from our programme, 
and the smaller ACS field obtained from the HST Data Archive, 
superimposed on an $R$ image from the Palomar Observatory 
Sky Survey. The image size is 11\,arcmin $\times$ 7.5\,arcmin. North is up, East 
to the left.}    
\label{fields}    
\end{figure}

The GMOS images correspond to two slightly overlapped fields 
(Fig.\,\ref{fields}), one of them centred on the galaxy 
NGC\,3610, and the other one located to the West (hereafter 
`N3610F' and `WestF', respectively). The point-sources located
in the overlapping region will allow any possible zero-point differences 
in the corresponding magnitudes to be determined. The exposure 
times were $4\times 450s$ in $g'$, $4\times 210s$ in $r'$, 
and $4\times 270s$ in $i'$. Each set of exposures was 
slightly dithered, in order to fill in the gaps of the GMOS 
field and to efficiently remove cosmic rays and bad pixels.

The ACS field (`ACSF') is also centred on NGC\,3610 
(see Fig.\,\ref{fields}). The exposure times were 
$6410s$ in filter $F555W$ (two $2330s$ observations plus a $1770s$ 
third one) and $6060s$ in filter $F814W$ (two $2330s$ observations 
plus a $1770s$ third one). The processed images were downloaded
from the HST Data Archive. \textsc{iraf} tasks
GEOMAP and GEOTRAN were used to register these images, due to 
differences in the position of their FOV.
Additionally, another ACS field of the 47\,Tuc outskirts was used 
to model the point-spread function (PSF). These observations were carried 
out in the same filters as the NGC\,3610 images, during June 2003 (programme 
9656), being the exposure time of $30s$.

\subsection{Photometry and point-source selection}
First, the surface-brightness profile of NGC\,3610 was obtained with the ELLIPSE 
and the corresponding synthetic galaxy, generated with BMODEL (both
\textsc{iraf} tasks), was subtracted from the original image in order to improve 
the detection of point-sources as much as possible. 

For the GMOS fields, the software SE\textsc{xtractor} \citep{ber96} 
was applied to the $i'$ image, which was selected because it has the higher 
signal-to-noise ratio ($S/N$), in order to obtain an initial point-source 
catalogue. As the effective radii ($R_{\rm eff}$) of classical GCs is usually 
smaller than 10\,pc \citep[e.g.][]{bru12}, at the adopted distance 
the NGC\,3610 GCs are seen as point-sources on our images. Then,
we used the SE\textsc{xtractor} parameter CLASS$\_$STAR to 
eliminate the extended sources from our catalogue. 
The photometry was performed with the DAOPHOT package \citep{ste87}
within \textsc{iraf}. A second-order variable PSF was generated 
for each filter, employing a sample of bright stars, well distributed
over the field. The final point-source selection was made with the 
$\chi$ and sharp parameters, from the task ALLSTAR.

In the case of ACS data, GC-like objects might be marginally resolved
at the distance of NGC\,3610 \citep[e.g.][]{cas13a,cas14}. Hence, we applied 
SE\textsc{xtractor} to images in both filters, but considered as likely 
GC candidates those sources with elongation smaller than 2 and FWHM 
smaller than 5\,pixel. Similar criteria have already been used for identifying GCs 
on ACS images \citep[e.g.][]{jor04,jor07}.      
Approximately 40 to 50 relatively isolated bright stars from the 47\,Tuc 
images were used to obtain the PSF for each filter. Then, aperture 
photometry was performed on the NGC\,3610 field, with an aperture radius of 
5\,pixel, which is almost three times the FWHM obtained for the foreground stars.
The software \textsc{ISHAPE} \citep{lar99} was used to calculate structural
parameters for the GC candidates, considering a typical $R_{\rm eff}$ of 0.35\,pixel 
(see Section\,\ref{GCrad}). Approximately 16\,GC and ultra-compact dwarf (UCD) candidates 
brighter than $F814W=23$\,mag, relatively isolated and with $0.3<R_{\rm eff}<0.4$\,pixel  
in both filters were used to calculate aperture corrections, resulting in 
$-0.07$\,mag for $F555W$ and $-0.12$\,mag for $F814W$. 
 
\subsection{Photometric calibration and background estimation}
A field of standard stars from the list of \citet{smi02} was observed 
for our GMOS programme, during the same nights than the field N3610F. 
We obtained the growth curve, and hence aperture corrections, from the standard 
stars' aperture photometry for several aperture radii. Then, we fit 
transformation equations of the form:

\begin{equation}
m_{std} = ZP + m_{inst} - K_{MK} \times (X-1) \nonumber
\end{equation}

\noindent where $m_{std}$ and $m_{inst}$ are the standard and 
instrumental magnitudes, respectively, and $ZP$ are the 
photometric zero-points for each band. $K_{MK}$ is the 
corresponding mean atmospheric extinction 
at Mauna Kea, obtained from the Gemini Observatory Web 
Page\footnote{http://www.gemini.edu/sciops/instruments/gmos/calibration}, 
and $X$ the airmass. For the $g'$ filter, part of the programme was
scheduled in February, while the rest of the observations were obtained
a month later, together with a photometric standards field. The zero-points 
for February and March observations were $ZP^{g'}=28.29\pm0.03$, 
and $ZP^{g'}=28.30\pm0.03$, respectively. The final assumed zero-points 
for the $g'$, $r'$, and $i'$ filters were $ZP^{g'}=28.30\pm0.03$, 
$ZP^{r'}=28.38\pm0.02$ and $ZP^{i'}=28.49\pm0.03$, respectively.
In the next step, we applied the Galactic extinction corrections obtained from 
\citet{sch11} to the calibrated magnitudes. Finally, considering the point-sources 
in common in both fields N3610F and WestF, we calculated the following zero-point 
differences, $\Delta_{g'}=-0.06$, $\Delta_{r'}=0.10$ and $\Delta_{i'}=0.13$.  
We applied these offsets to the WestF catalogue and referred 
the photometry to the field N3610F.

In the case of ACS data, the calibrated magnitudes were obtained using
the relation:

\begin{equation}
m_{std} = m_{inst} + ZP \nonumber
\end{equation}

\noindent for each filter, with zero-points $ZP_{F555}= 25.724$ and $ZP_{F814}= 25.501$, 
taken from \citet{sir05}, so that the resulting calibrated magnitudes correspond  
to $V$ and $I$ filters, respectively.

The ACS data has already been used to study the properties of the GC candidates by \citet{gou07}, 
who applied along their work corrections for contamination by background objects. Such contamination 
was calculated as the compact objects detected beyond a galactocentric radius (hereafter designated 
with `$R_g$') $R_g =$100\,arcsec (i.e. 1.67\,arcmin), which they considered might be slightly overestimated.
In fact, thanks to the larger area covered by our GMOS data, we will show in the following 
that GCs are present up to $R_g =$ 4\,arcmin. Thus, it is worth doing a new analysis of the ACS data, 
together with the new ones from GMOS, but taking into account a more precise estimation of the 
correction for background contamination. 

In order to estimate the background contamination for the GMOS data, we considered the 
point-sources within an area of $16.5\,{\rm arcmin}^2$ at the Western side of field WestF, 
located at $\sim 5$\,arcmin from the centre of NGC\,3610. It will be identified 
in the following as the `comparison region'. As we lack an appropriate comparison field 
for the ACS data, we will use the corrections obtained from the comparison region of the 
GMOS data, taking into account the different depths and completeness corrections of 
both photometries.

\subsection{Completeness analysis}
\begin{figure}    
\includegraphics[width=85mm]{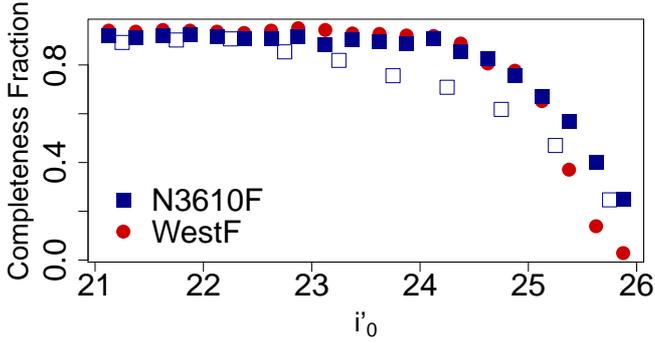}    
\caption{Completeness curves for the two GMOS fields, as function of $i'_0$ 
magnitude. The bin width is 0.25\,mag. For the field N3610F, we distinguished  
between artificial 
stars at less than 1\,arcmin from NGC\,3610 centre (open squares), and further 
than this limit (filled squares).}    
\label{comp}    
\end{figure}

In order to estimate the photometric completeness for our GMOS fields,
we added 250 artificial stars to the images of the three filters,  
distributed over the entire fields in an homogeneous way. Their colours span the 
expected ranges for GCs and magnitudes equally generated
for $21 < i'_0 <26$. As this process was repeated 40 times, 
we achieved a whole sample of 10\,000 artificial stars. 
Their photometry was performed following exactly the same steps as for the science 
images. After selecting the definite point-sources, we obtained the completeness 
curves shown in Fig.\,\ref{comp}. 
For the field N3610F, we discriminated between artificial stars located at less 
than 1\,arcmin from the NGC\,3610 centre (open squares), and further than this 
limit (filled squares). The completeness 
functions for the outer region of N3610F and the field WestF are very similar, 
achieving the 70 per cent at $i'_0\sim 25$. This value has been considered as the 
faint magnitude limit in the following analysis.

\begin{figure}    
\includegraphics[width=85mm]{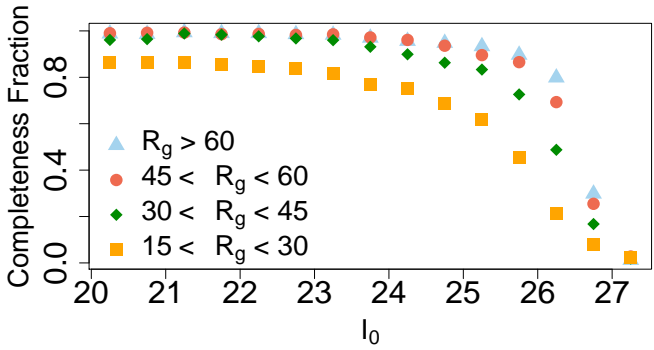}    
\caption{Completeness curves for the ACSF field, as a function 
of $I'_0$ magnitude, for different galactocentric ranges ($R_g$ in arcsec). The 
bin width is 0.25\,mag.}    
\label{comp2}    
\end{figure}    

In the case of the ACSF field, we added 50 artificial stars per image, spanning 
the colour range of GCs and $21.5 < I_0 <27$. We repeated the process to
achieve a final sample of 50\,000 artificial stars, which allowed us to 
calculate the completeness curves for different galactocentric ranges 
(Fig.\,\ref{comp2}). As well as for the GMOS fields, the 
photometry was developed in the same manner as for the science field. 
We selected as magnitude limit $I_0 \sim 25.5$, which might represent a 
conservative selection for the outer galactocentric ranges, but 
corresponds to $\sim 60$ per cent in the range 15\,arcsec $< R_g < $ 30\,arcsec. 
For the GMOS and ACS fields, 
it is clear the decrease of the completeness as we approach the galaxy centre. 

\section{Globular cluster system of NGC\,3610}

\subsection{Selection of GC candidates}
\label{dcmsec}
\begin{figure}    
\includegraphics[width=85mm]{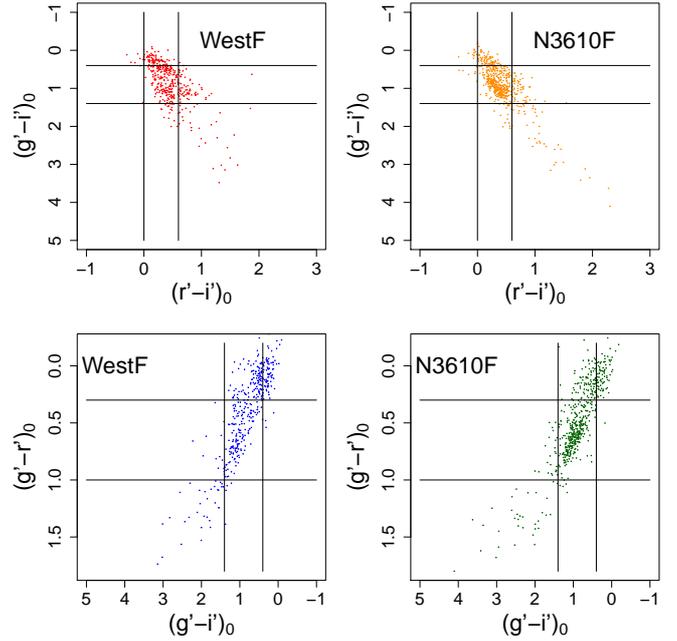}    
\caption{Colour-colour diagrams for both GMOS science fields. Solid lines 
indicate the colour ranges of selected GC candidates (see Section\,3.1).}   
\label{dcc}    
\end{figure}

The colour-colour diagrams $(g'-i')_0$ versus $(r'-i')_0$ and $(g'-r')_0$
versus $(g'-i')_0$ for all the point-sources in both GMOS fields are presented 
in Fig.\,\ref{dcc}. As the colours of {\it bona-fide} GCs fall within narrow ranges, 
we can use these diagrams to distinguish them from the contamination  
\citep[e.g.][ and references therein]{cas15a}. We selected as GC candidates 
the point-sources in the following ranges: $0.4 < (g'-i')_0 < 1.4$, 
$0.3 < (g'-r')_0 < 1$ and $0 < (r'-i')_0 < 0.5$. These adopted limiting colours 
are similar to those chosen by \citet{cas15a} and \citet{esc15}.

\citet{str03,str04b} inferred ages and metallicities for a sample of 13 GC 
candidates in NGC\,3610, by means of Lick/IDS indices, resulting in a 
majority of old GCs. However, two GC candidates in their sample turned out 
to be young and very metal-rich ($\sim 2$\,Gyr and $[Z/H] \sim 0.5$). 
In order to confirm that even these GCs would be identified with the present 
selection criteria, we obtained the expected colours in our bandpasses from 
the theoretical models of single stellar populations (SSP) by 
\citet{bre12}, using their web-based tool 
\footnote{http://stev.oapd.inaf.it/cgi-bin/cmd}. If we consider a \citet{cha01}
lognormal IMF and the reddening corrections applied to our point-sources 
catalogue, a SSP with these ages and metallicities would have  
$(g'-i')_0 \sim 1$, $(g'-r')_0 \sim 0.7$ and $(r'-i')_0 \sim 0.3$,
which fulfill our colour criteria.

\begin{figure}    
\includegraphics[width=85mm]{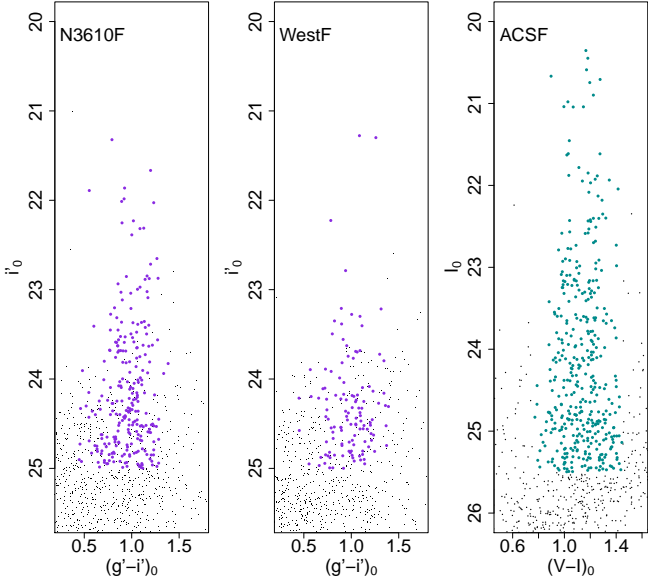}    
\caption{Colour-magnitude diagrams for the two GMOS fields and the ACS
field. Filled circles highlight those objects that fulfill the colours 
and magnitude criteria applied to select GC candidates.}    
\label{dcm}    
\end{figure}

In the ACS photometry, we selected as GC candidates the point-sources or 
marginally resolved sources with colours ranging $0.75 < (V-I)_0 < 1.4$ \citep{bas08}. 
In Fig.\,\ref{dcm}, the left-hand side and middle panels show the $i'_0$ 
versus $(g'-i')_0$ colour-magnitude diagrams (CMD) for the two GMOS fields. The 
black dots represent the objects that fulfill the point-source criteria, after 
the $\chi$ and sharp selection. The filled circles identify those that 
fulfill the GC candidates' colours and magnitude criteria indicated before.
The right-hand side panel 
shows the $I_0$ versus $(V-I)_0$ CMD for the ACS field, with GC candidates 
highlighted with filled circles.

\begin{figure}    
\includegraphics[width=80mm]{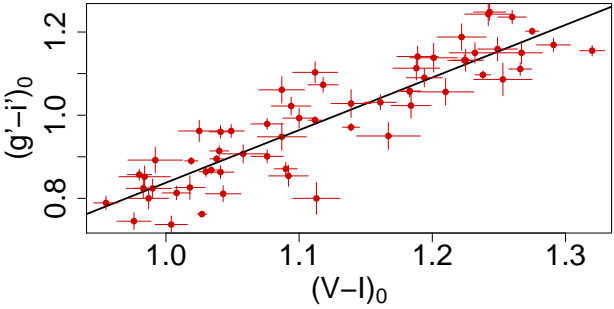}    
\caption{$(g'-i')_0$ versus $(V-I)_0$ colours for a sample of bright GC 
candidates with available magnitudes in both photometric systems.}    
\label{vigi}    
\end{figure}

In order to calculate the $(V,I)$ to $(g',i')$ transformation, we selected 
the 62 GC candidates with available colours in both photometric systems, with 
$i'_0 < 24$\,mag. Fig.\,\ref{vigi} shows $(g'-i')_0$ versus 
$(V-I)_0$ colours for these objects. The solid line represents the 
transformation obtained with a linear least-squares fit, 

\begin{equation}
(g'-i')_0 = 1.26\pm0.06 \times (V-I)_0 - 0.43\pm0.07
\end{equation}

\noindent As a comparison, from an equivalent transformation 
presented by \citet{fai11}, their equation\,3, we can 
derive the coefficients 1.25 and 0.4, respectively. 
According to this transformation, we improve our GC colour limits 
as $0.66 < (V-I)_0 < 1.45$ so that they agree with the ones adopted 
for $(g'-i')_0$. 
Between $I_0$ and $i'_0$ there is only a zero-point difference, 

\begin{equation}
i'_0 = I_0 + 0.43\pm0.01
\end{equation}

\subsection{Colour distribution}

\begin{figure}    
\includegraphics[width=85mm]{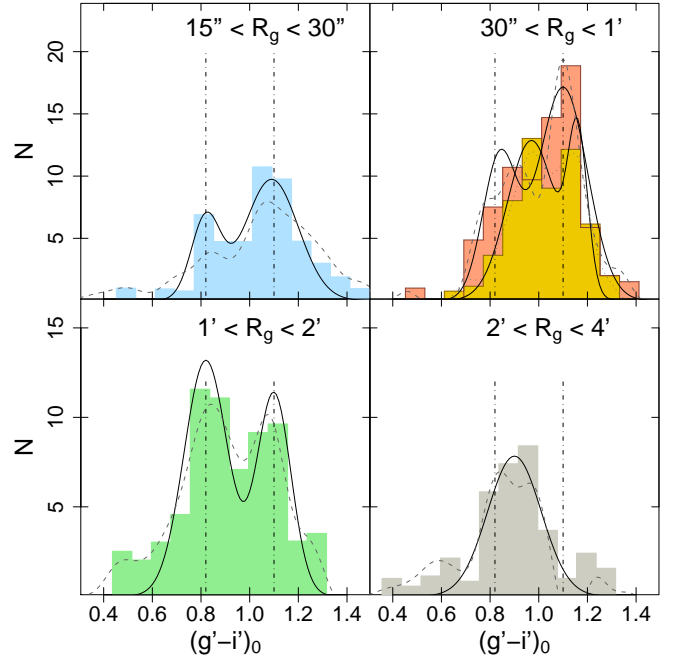}    
\caption{ Background-corrected colour distribution (dashed lines) for the GC candidates 
from ACS photometry (upper left panel), overlapping ACS and GMOS data (upper right panel, 
GMOS sample is the smaller one), 
and GMOS photometry (lower panels), for four radial intervals as indicated in the respective 
panels. Solid lines show the results from GMM fitting and vertical dot-dashed lines 
indicate the typical colours of blue and red peaks ($(g'-i')_0 =$\,0.82 and 1.10, respectively). 
Please, notice the different scales in upper and lower panels.}   
\label{dcol}    
\end{figure}

In order to analyse the GC colour distribution, we separated the sample in four 
radial regimes, corresponding to 15~$ < R_g <$ 30\,arcsec, 30\,arcsec $ < R_g < $ 1\,arcmin, 
1~$ < R_g <$ 2\,arcmin and  2~$< R_g <$ 4\,arcmin. To this aim, we combined 
GMOS and ACS data, transforming ACS photometry into $(g'-i')$ colour and $i'_0$ 
magnitude by means of equations\,(1) and (2). Both data sets will overlap in only one radial range 
(30\,arcsec $ < R_g < $ 1\,arcmin). 
Fig.\,\ref{dcol} shows the background-corrected $(g'-i')$ colour distribution (dashed lines)
for GC candidates in these four different radial ranges, with $i'_0< 25$ and applying a bin 
width of 0.08\,mag. 

The same procedure was applied in the four radial ranges. First, we obtained a clean sample 
of GC candidates, randomly selecting sources from the comparison region and deleting GC candidates 
with similar $(g'-i')$ colours, until we reached the expected number of objects due to contamination. 
Then, we applied to each sample the Gaussian Mixture Modeling test \citep[{\textsc GMM},][]{mur10} 
to determine whether their respective colour distributions are likely to be represented by the sum 
of two Gaussians.
For each case, this procedure was repeated 25 times, and the mean results are listed in Table~\ref{fit}, 
including the mean colours, dispersions, and the fractions ($f$) for each sub-population. 
The two last columns correspond to the $DD$ parameter, which is related to the separation of the mean 
values  
and indicates whether an specific bimodal distribution provides a realistic fit (a meaningful 
bimodal case is accepted for $DD > 2$), and the kurtosis ($\kappa$), which is 
expected to be negative for bimodal distributions. For the sake for comparison, we have included 
in Table~\ref{fit} the results of the unimodal fits even in the cases where $DD$ and $\kappa$ pointed 
to a bimodal fit. According to the calculated $DD$ and $\kappa$ parameters, the three inner subsamples seem 
to be better described by bimodal distributions, 
but just a single Gaussian should be fitted to the outermost one. Moreover,
it is also noticeable from Fig.\,\ref{dcol} how different the GC colour distributions
are when we discriminate them according to $R_g$. 

For the inner radial subsample, 15 $< R_g <$ 30\,arcsec, the colour distribution 
is better fitted by two Gaussians, with expected values for the blue and red peaks 
(mean colours), i.e. $(g'-i')_0 \sim$~0.8 and 1.1 \citep[e.g.][]{esc15}. This distribution 
is dominated by the red GC sub-population that represents more than 70\,per cent of the subsample.  
The GC distribution for 30\,arcsec $< R_g <$ 1\,arcmin is also better fitted by two Gaussians, 
as indicated by the $DD$ and $\kappa$ parameters for both ACS and GMOS data. 
Both distributions are shown in Fig.\,\ref{dcol} (upper right panel) and the histograms look similar, with 
a dominant red sub-population and a few more GCs in the ACS sample. However, the fits are not 
so similar, as only the ACS data have blue and red peaks in agreement with the ones estimated for 
the previous (innermost) radial range. The colour of the bluer peak estimated for the GMOS subsample 
is in the middle between the usual blue and red peaks, though it must be taken into account that 
the differences may be due to few GCs. Anyway, the colour distribution for both subsamples of ACS 
data agrees with that for the inner subsample of GMOS data, in the sense that they are all dominated 
by red GCs. This is also evident from the radial projected distribution depicted 
in Fig.\,\ref{drad} (middle panel), where there is a clear deficiency of blue GCs in the 
inner region as compared to the profile for larger $R_g$. 

The intermediate radial GMOS subsample (lower left panel) shows the more clear bimodal distribution, 
with blue and red peaks in the expected colours (indicated in all panels with dot-dashed vertical lines 
to facilitate comparison). The fractions of blue and red clusters 
are 60 and 40 per cent, respectively, with the blue sub-population presenting an extension 
towards very blue colours. Finally, the outermost GMOS subsample (lower right panel) is quite 
anomalous as a unimodal distribution is present, with a mean $(g'-i')_0 = 0.87$. We are aware that 
this is a small subsample, where a bunch of blue clusters also seems to be present, but the group is 
dominated by GCs of `intermediate' colours. Their mean colour agrees with the bluer peak obtained 
for the inner GMOS subsample (30\,arcsec $< R_g <$ 1\,arcmin), so they may be probable present 
at smaller radii too. We have already detected GCs of `intermediate' colours in other galaxies 
that experienced recent mergers, and will come back to this in the Discussion.

\begin{table*}
\begin{center}
\caption{Parameters of the {\textsc GMM} fitting to the colour distribution, 
considering different radial ranges. $\mu_j$, $\sigma_j$ and $f_j$ correspond 
to the mean, dispersion and fraction for each Gaussian component ($j=1$: blue GCs 
and $j=2$: red GCs). We refer to the text for the meanings of $DD$, and $\kappa$.}
\label{fit}
\resizebox{\textwidth}{!}{   
\begin{tabular}{@{}ccccccccc@{}}   
\hline   
\multicolumn{1}{@{}c}{}&\multicolumn{1}{c}{$\mu_{1}$}&\multicolumn{1}{c}{$\sigma_{1}$}&\multicolumn{1}{c}{$f_{1}$}&
\multicolumn{1}{c}{$\mu_{2}$}&\multicolumn{1}{c}{$\sigma_{2}$}&\multicolumn{1}{c}{$f_{2}$}&\multicolumn{1}{c}{DD}&\multicolumn{1}{c}{$\kappa$}\\   
\hline   
\multicolumn{9}{c}{ACS data}\\   
\hline   
$15'' < R_g < 30''$ & & & & & & & $3.19\pm0.46$ & -0.67 \\
Unimodal & $1.03\pm0.02$ & $0.15\pm0.01$ & $-$ & $-$ & $-$ &$-$ &  & \\
Bimodal & $0.82\pm0.03$ & $0.06\pm0.02$ & $0.27\pm0.09$ & $1.09\pm0.03$ & $0.11\pm0.02$ &$0.73\pm0.09$ &  & \\
$30'' < R_g < 1'$ & & & & & & & $2.89\pm0.45$ & -0.81 \\
Unimodal & $1.02\pm0.02$ & $0.15\pm0.01$ & $-$ & $-$ & $-$ &$-$ &  & \\
Bimodal & $0.84\pm0.03$ & $0.07\pm0.02$ & $0.32\pm0.11$ & $1.10\pm0.02$ & $0.10\pm0.02$ &$0.68\pm0.11$ &  & \\
\hline   
\multicolumn{9}{c}{GMOS data}\\   
\hline    
$30'' < R_g < 1'$ & & & & & & & $2.15\pm0.17$ & -0.55 \\
Unimodal & $1.03\pm0.01$ & $0.13\pm0.01$ & $-$ & $-$ & $-$ & $-$ &  &  \\
Bimodal & $0.97\pm0.03$ & $0.10\pm0.01$ & $0.72\pm0.05$ & $1.16\pm0.02$ & $0.04\pm0.01$ &$0.28\pm0.05$ &  & \\
$1' < R_g < 2'$ & & & & & & & $3.15\pm0.13$ & -0.97 \\
Unimodal & $0.93\pm0.01$ & $0.16\pm0.01$ & $-$ & $-$ & $-$ & $-$ &  &  \\
Bimodal & $0.82\pm0.02$ & $0.09\pm0.01$ & $0.60\pm0.03$ & $1.10\pm0.01$ & $0.07\pm0.01$ &$0.40\pm0.03$ &  & \\
$2' < R_g < 4'$ & & & & & & & $1.82\pm0.20$ &  \\
Unimodal & $0.87\pm0.02$ & $0.11\pm0.01$ & $-$ & $-$ & $-$ & $-$ &  & \\
\hline
\end{tabular}  
}
\end{center}
\end{table*}   

Summarising, thanks to the larger FOV of GMOS we have been abe to study the GC colour distribution 
over the whole radial extent, reaching the outer sub-populations never analysed before. We recovered 
the dominant inner red GC sub-population in agreement with the results from the ACS data, and found 
a bimodal GC distribution 
at intermediate radii as well as a small number of clusters with `intermediate' colours in the outskirts, 
though they may be present at inner regions too. 

\subsection{Projected spatial and radial distributions}
\label{projdist}
\begin{figure}    
\includegraphics[width=80mm]{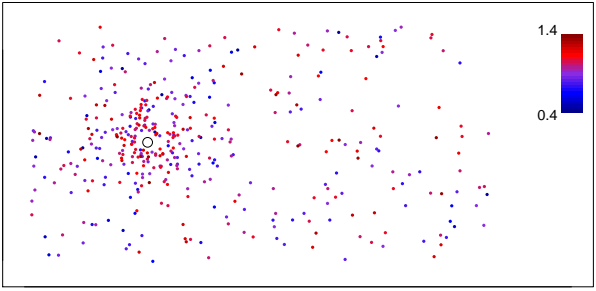}    
\caption{Projected spatial distribution for GC candidates from ACS and GMOS 
data. The centre of the galaxy is highlighted with an open circle. 
North is up, East to the left. The field of view is 
$10.2\times5.5\,{\rm arcmin}^2$. The colour bar depicted on the top 
right-hand side corresponds to $(g'-i')_0$.}    
\label{espa}    
\end{figure}

The projected spatial distribution for GC candidates in both GMOS fields, combined 
with those selected from the ACS data in the inner region, is 
shown in Fig.\,\ref{espa}. The colour range, as depicted in the 
Figure, spans $0.4< (g'-i')_0 <1.4$ and the centre of NGC\,3610 is indicated 
with an open circle. 
It can be seen that red GCs are more concentrated towards the galaxy, while 
clusters of bluer colours appear to dominate at larger radii. 

\begin{figure}    
\includegraphics[width=85mm]{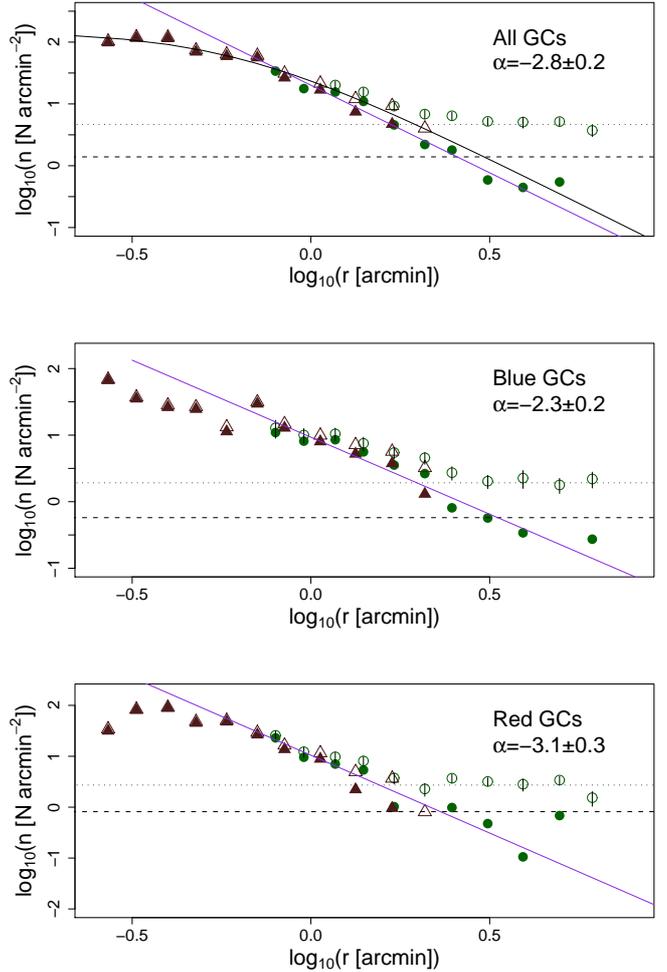}    
\caption{Raw and background corrected radial distributions for GC 
candidates (open and filled circles, respectively). Triangles and 
circles identify the combined data, from ACS and GMOS, respectively. 
The solid straight lines  
represent the respective power-laws, fitted by least-squares, and the 
respective slopes $\alpha$ are depicted in each panel. 
The dashed lines correspond to the respective background levels, 
while the dotted ones indicate 30 per cent of the background level, 
used to define the GCS extension. The solid curved line in the upper 
panel shows the fit of a Hubble profile (see text for the results). 
Please, notice the different scale in the lower panel.}
\label{drad}    
\end{figure}

Fig.\,\ref{drad} presents the raw and background-corrected radial 
distributions for the GC candidates from GMOS data (open and filled circles, 
respectively), and corrected by completeness. The open and filled triangles  
represent the analogous distributions, but from ACS data. Upper panel corresponds  
to the whole GC sample, while middle and lower panels present similar plots but 
for blue and red GCs, respectively. It can be seen that results from GMOS and ACS 
data are in good agreement within the range of overlapping $R_g$. Here, we do not consider 
separately the intermediate colour clusters (located in the outer region) as they are too few 
to derive a radial profile.   
The adopted colour limit between both GC sub-populations is $(g'-i')_0=0.95$ \citep{fai11}, 
which corresponds to $(V-I)_0=1.1$. 
The respective background levels are indicated with dashed lines, and were determined from 
the comparison region described in Section\,2.3.

The background-corrected radial profiles were fitted by power-laws with slopes 
$-2.8\pm0.2$, $-2.3\pm0.2$ and $-3.1\pm0.3$ for the entire sample, blue and red GCs, 
respectively (indicated with the solid lines). Due to incompleteness in the central region, 
the fits were performed for radii $0.5 < r < 4$\,arcmin. These results 
clearly confirm that the red sub-population is more concentrated towards the 
galaxy while the bluer clusters one extends further away.
 
We established the GCS extent as the galactocentric distance at 
which the background-corrected surface density is equal to the 
30 per cent of the background level (dotted lines). This criterion has 
been previously used in similar studies \citep[e.g.][]{bas06a}, 
and implies a new determination of the GCS extent of $\sim 4$\,arcmin, 
i.e., $\sim 40$\,kpc at the adopted distance for NGC\,3610. 

For both GC sub-populations, the radial density profiles flatten towards the 
galaxy centre. As the completeness analysis was performed for  
different galactocentric radii, in order to take into account the effect of 
the galaxy surface-brightness profile, the flattened 
radial profiles might imply a real paucity of GCs in the
inner regions of the galaxy. In order to consider the slope 
change, we fitted to the radial profile of the entire sample 
($\rm r < 4$\,arcmin) 
a Hubble profile \citep{bin87,dir03a} of the form: 

\begin{equation}
n(r) = a \left( 1+ \left(\frac{r}{r_0} \right)^2 \right)^b
\end{equation}

\noindent where $a = 152\pm13$, $r_0 = 0.62\pm0.05$, and $b = -1.42\pm0.08$ 
(solid curved line in Fig.\,\ref{drad}, upper panel).  
This profile provides a much better fit to the inner radial distribution.
As well as GCs formation requires specific environmental conditions and 
merger remnants are the places where YMC (young massive clusters) are found  
\citep[e.g.][]{kru14}, these conditions also favour their tidal disruption 
\citep{kru12,kru15}. 
\citet{broc14} also points to the relevance of the erosion of GCs in the
present-day characteristics of its GCS. The erosion might be responsible 
for the evolution of GCs mass function from the initial power-law to a 
bell-shaped distribution. The GCs destruction mainly occurs up to the
galaxy half-light radius, with smaller upper limit radius 
and efficiency when we move towards more massive and extended elliptical 
galaxies. The degree of the initial radial anisotropy in the velocity 
distribution of the GCS also play a main role in the fraction of eroded 
GCs and the radius at which it is efficient \citep{broc14}.

\subsection{Luminosity function and GC population}

\begin{figure}    
\includegraphics[width=85mm]{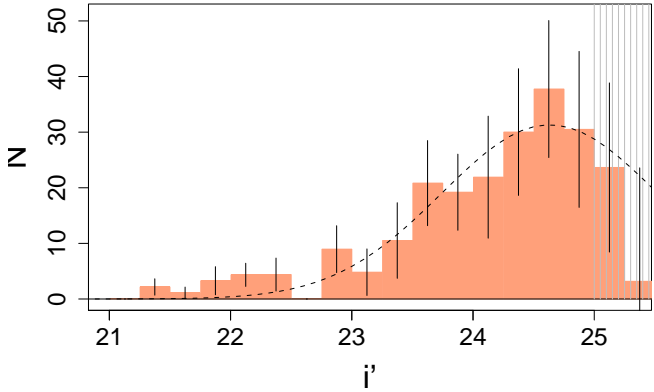}    
\caption{Background and completeness corrected luminosity 
function (GCLF) for GCs based on the GMOS data. The errorbars 
assume Poisson uncertainties for science and background 
measurements, and the bin width is 0.25. The vertical lines 
indicate the luminosity range fainter than $i'_0=25$, that has 
not been included in the GCLF fitting. 
}    
\label{lf}    
\end{figure}

Fig.\,\ref{lf} shows the background and completeness corrected 
GCLF for the GC candidates selected from the GMOS data. The errorbars 
assume Poisson uncertainties for science and background measurements,
and the bin width is 0.25\,mag. The bins filled with vertical grey lines 
have not been considered in the GCLF fitting, due to the declining 
completeness (i.e. according the limit $i'_0=25$ adopted in Section\,2.4).
The turn-over magnitude (TOM) and dispersion obtained from the 
least-squares fit of a Gaussian model are $i'_{0,TOM}=24.6\pm0.25$ 
and $\sigma=0.9\pm0.24$.

Old GC populations in early-type galaxies usually present a 
Gaussian GCLF, with a TOM in the $V$-band of $M_{\rm V_ {\rm TOM}}\sim-7.4$  
\citep[e.g.][]{ric03,jor07}, denoting the universality of the GCLF 
that is a reliable distance indicator. 
On the basis of ACS data, \citet{gou07} found 
that the GCLF for red GCs in NGC\,3610 deviates from the usual Gaussian 
distribution and can be fitted by a power-law. It has been clearly shown  
that power-law luminosity functions correspond to young stellar clusters, 
not {\it bona-fide} old GCs \citep[e.g.][]{whi14}. Our GMOS photometry is 
not as deep as the ACS one, so we cannot compare directly the behaviour 
of both GCLFs. In addition, if we want to use the ACS photometry that we 
have re-done, the background correction to be applied to these data was 
obtained from the GMOS images, so the same limiting magnitude is valid.  

We have adopted for NGC\,3610 the SBF distance modulus ($m-M \sim 32.7$), 
which implies a `standard' $V_{\rm TOM} \sim 25.3$ according to the 
$M_{\rm V_ {\rm TOM}}$ quoted above. If we use the 
colour and magnitude transformations derived in Section\,3.1 and consider for the 
GC candidates a mean colour $(g'-i')\sim 0.96$ (estimated from a clean sample), 
the resulting TOM is $i'_0 \sim 24.6$, in agreement with the value obtained directly 
from the data. As we are analysing the global GCLF, and not those of blue and red GCs separately, 
the TOM from the fitted Gaussian agrees with what would be a `standard' TOM, according to 
the SBF distance. We understand this result as just a consequence that we are considering 
the whole GC sample, which is dominated by truly `old' GCs, and we are not able to study 
separately the LF of the intermediate-age GC sub-population.  

In order to estimate the total GC population, with the advantage of having a large FOV 
with the GMOS data, we first consider the number of GCs obtained from the integration of 
the Hubble radial profile. Then, we apply a correction to take into account that we are covering 
a fraction of the fiducial Gaussian adopted as GCLF, up to our magnitude limit $i'_0 \sim 25$.
With this procedure, the estimated `old' GC population of NGC\,3610 results $500\pm110$ members. 

Assuming the total $V_0$ magnitude of NGC\,3610, obtained from NED, 
as $V_{\rm tot_0}= 10.7\pm0.13$, 
and the distance modulus $m-M =32.7\pm0.1$, the absolute V magnitude of 
the galaxy is $M_V=-22.0\pm0.16$. Then, the specific frequency, i.e. 
the number of GCs per unit galaxy luminosity \citep{har81}, 
for the old GC population results $S_N=0.8 \pm 0.4$. It is approximately in agreement,  
within the errors, with the specific frequencies obtained by \cite{whi97} ($S_N=0.6 \pm 0.14$) and 
 \citet{gou07} ($S_N=1.4 \pm 0.6$). 
Such values correspond 
to the lower limit of the range of $S_N$ obtained for elliptical galaxies of 
similar luminosity \citep{har13}. In fact, the most frequent galaxy type with 
$S_N < 1$, in the brightness range corresponding to massive galaxies ($M_V < -20$), 
are not ellipticals but spirals (Harris et al., their fig.~10). 

\subsection{Effective radii of globular clusters}   
\label{GCrad}
\begin{figure}    
\includegraphics[width=42mm]{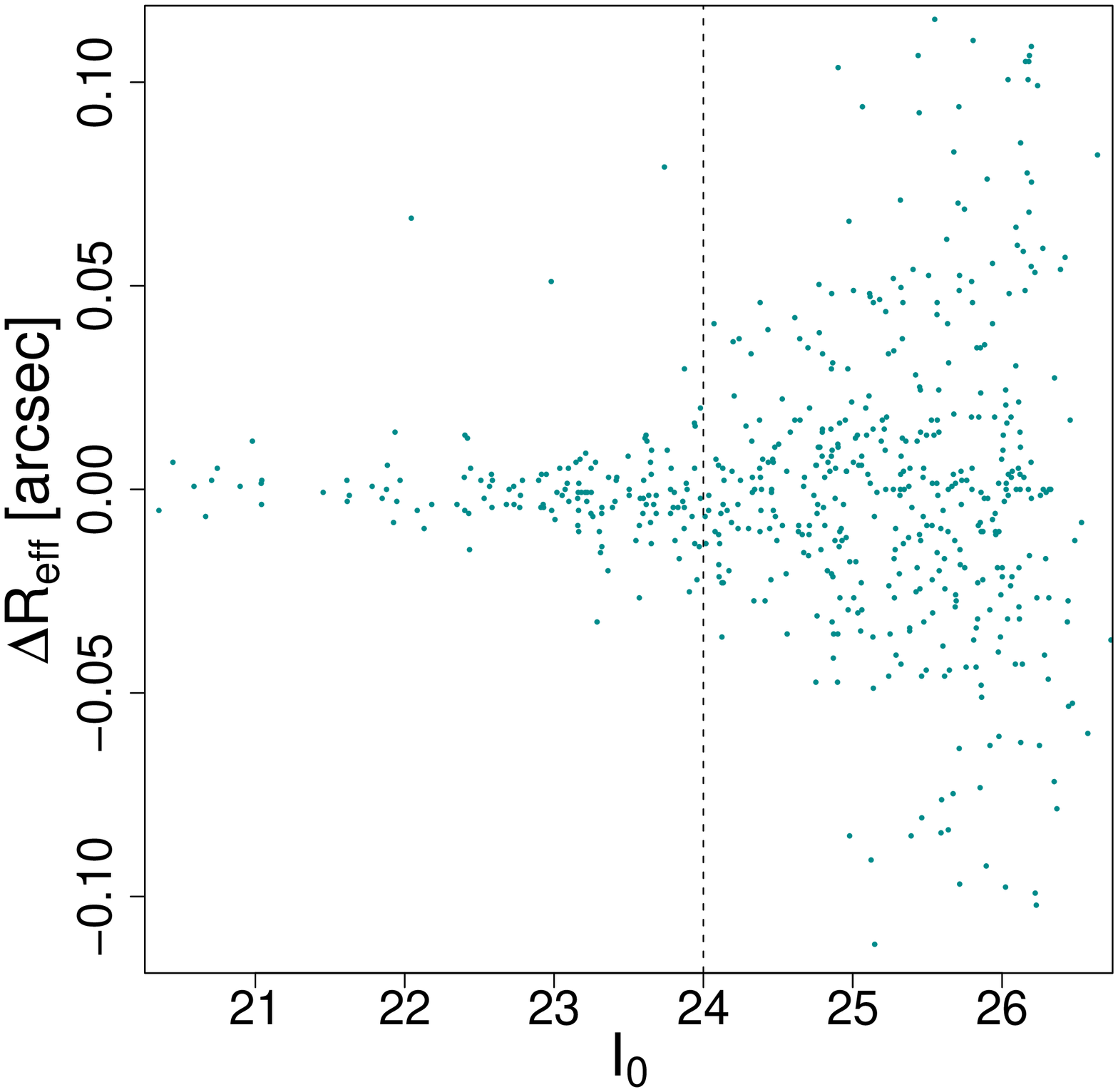}
\includegraphics[width=42mm]{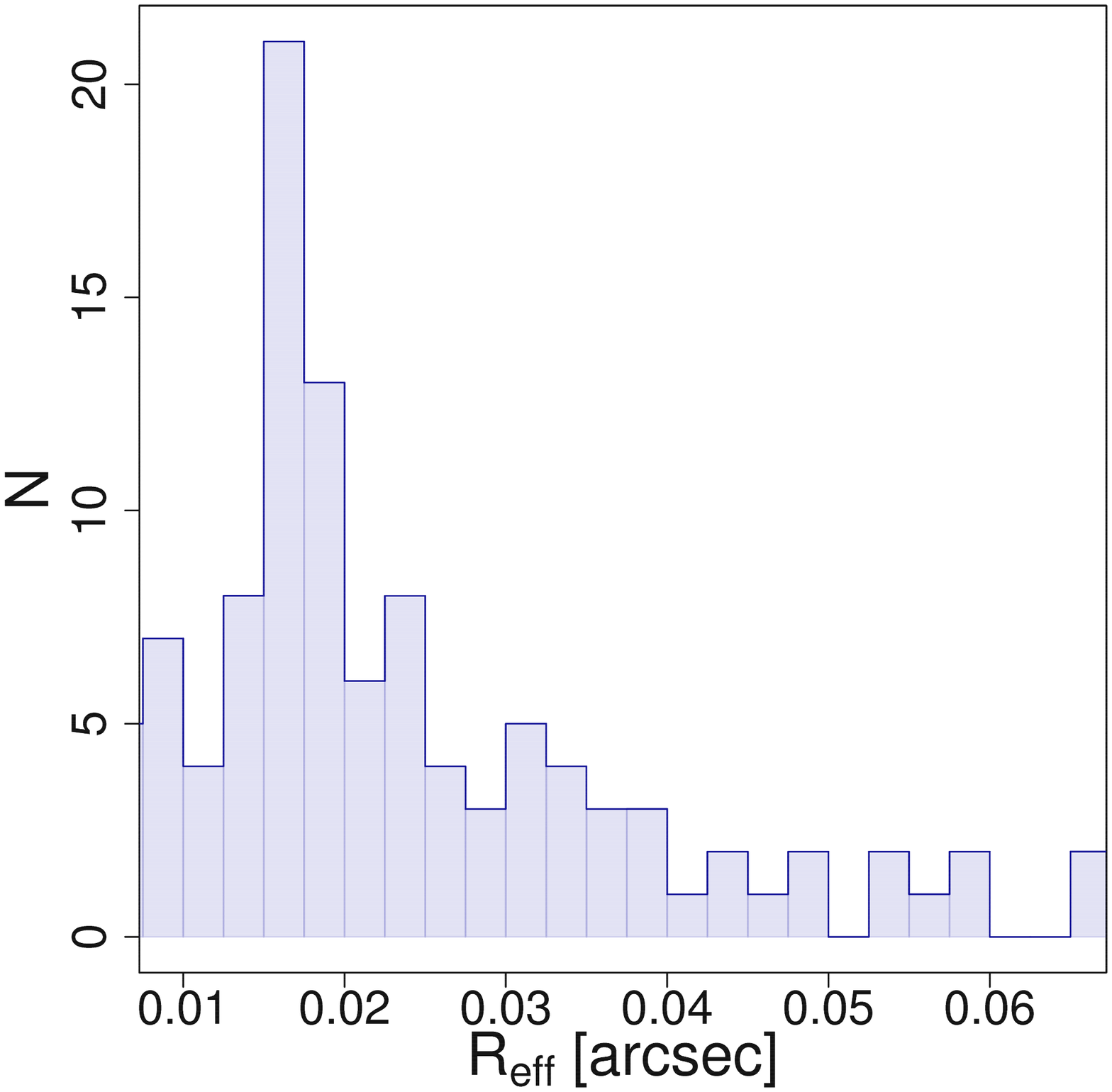}
\caption{Left: difference between $R_{\rm eff}$ measured in $V$ and $I$  
filters for GC candidates from ACS data, in arcsec. The vertical 
dashed line represents the magnitude limit up to which $R_{\rm eff}$ is reliable.
Right: $R_{\rm eff}$ distribution for GC candidates brighter than $I_0 = 24$.}    
\label{dreff}    
\end{figure}

The size of GCs, measured as their effective ($R_{\rm eff}$) or half-light radius, 
is one of the parameters that characterise these stellar systems and helps our understanding 
of their formation and evolution \citep [e.g.][and references therein]{puz14}. 
As a consequence of the outstanding resolution of the ACS data, it is 
possible to compute $R_{\rm eff}$ of GCs up to few parsecs, at 
similar distances than NGC\,3610 \citep[e.g.][]{cas13a,cas14}. For this purpose, 
we used the software \textsc{ISHAPE} \citep{lar99}, designed to
calculate structural parameters for marginally resolved objects by fitting their 
surface-brightness profiles with analytical models, convolved with a PSF. As 
mentioned in Section\,2.2, for each filter we derived the PSF from
observations of 47\,Tuc outskirts, with images obtained during the same month 
as those of NGC\,3610. We selected a King30-profile, i.e. a King profile with 
concentration parameter c = 30, being $c$ the ratio of tidal over core 
radius \citep{kin62,kin66}. 
We did not consider any possible eccentricity, but large values
are not expected for GCs nor for UCDs \citep[e.g.][]{har09,chib11}.
The left-hand panel of Fig.\,\ref{dreff} shows the difference between the 
$R_{\rm eff}$ measured in $V$ and $I$ for the GC candidates as a function 
of $I_0$, which presents an even distribution around zero. The mean $R_{\rm eff}$ 
between those obtained in both filters will be adopted as the final values.  
\citet{lar99} indicated that structural parameters calculated by
\textsc{ISHAPE} are reliable when the object $R_{\rm eff}$ is at least
one tenth of the PSF, for $S/N \sim 50$ or higher. This latter
condition is fulfilled in both filters for GCs brighter than $I_0 = 24$, 
indicated with a vertical dashed line in the Figure. In both filters, the 
derived PSF had  
$FWHM \sim 0.08$\,arcsec, which implies for the King30-profile 
$R_{\rm eff} \sim 0.12$\,arcsec. 
Hence, reliable $R_{\rm eff}$ measurements were obtained for GCs brighter 
than $I_0 = 24$ and with $R_{\rm eff} > 0.01$\,arcsec. This agrees with
the results, as small differences in the $R_{\rm eff}$ calculated with both 
filters were obtained for GCs brighter than the quoted magnitude limit.

\begin{figure*}    
\includegraphics[width=140mm]{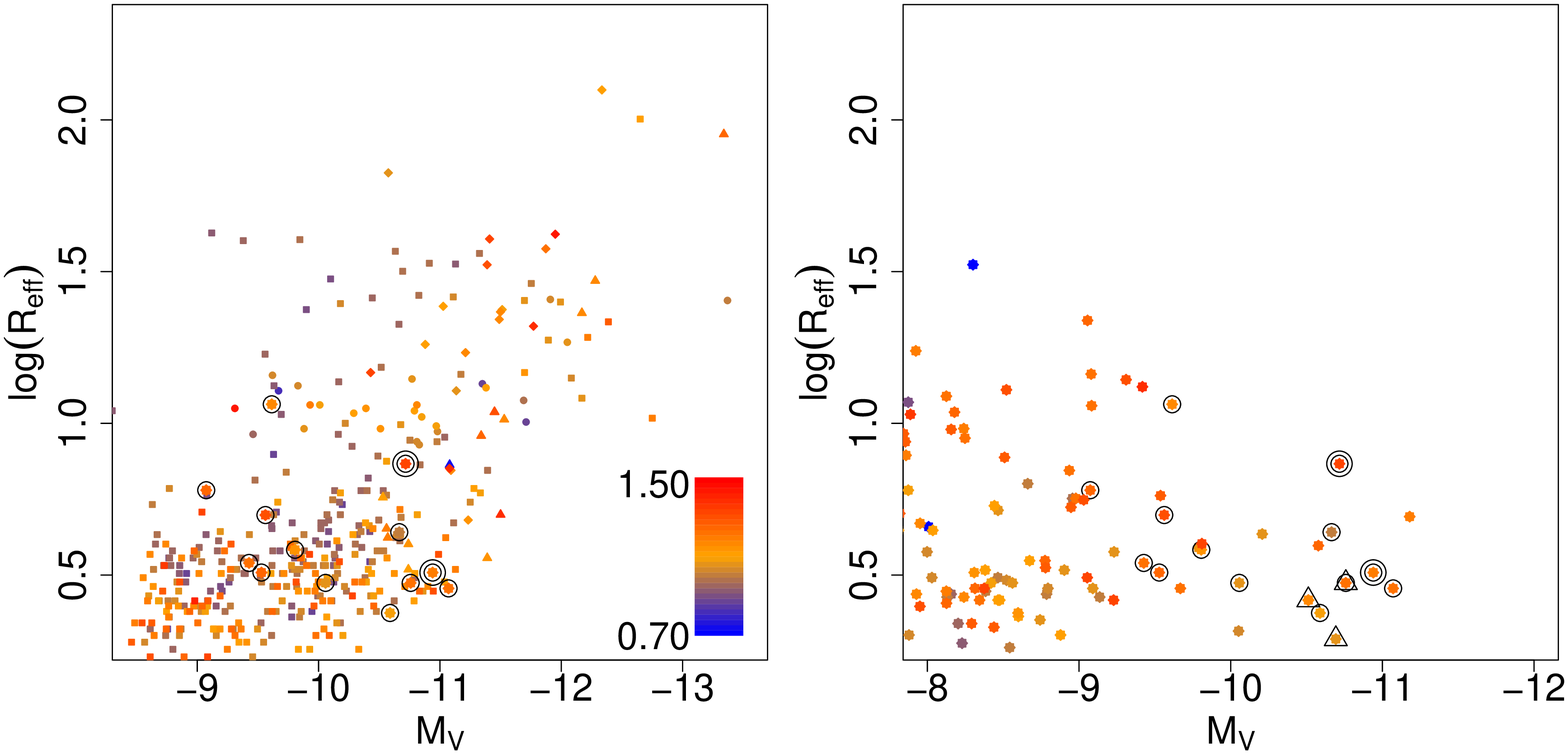}
\caption{Left: $R_{\rm eff}$[pc] versus $M_V$ for NGC\,3610 spectroscopically 
confirmed GCs \citep[][highlighted with open circles]{str03,str04b}, 
together with GCs/UCDs from Virgo \citep{bro11}, Fornax \citep{mie04}, 
Antlia \citep{cas13a,cas14}, Hydra \citep{mis11}, and Coma \citep{chib11}. The two 
clusters identified with double circles correspond to W6 and W11, the young and metal-rich 
ones from the Strader sample. Right: Similar plot for NGC\,3610 GC candidates, including the same 
spectroscopically confirmed GCs (open circles), and three GC-like objects 
with available $R_{\rm eff}$ measurements (triangles), whose positions match with 
X-ray point-sources \citet{liu11}. The colour bar represents the $(V-I)_0$ colours of GCs and UCDs  
in both panels.
}    
\label{mvreff}    
\end{figure*}

The right-hand panel of Fig.\,\ref{dreff} shows the $R_{\rm eff}$ distribution
for GC candidates brighter than $I_0 = 24$. 
The binwidth is $0.0025$\,arcsec. The distribution maximum is about 
$0.015-0.0175$\,arcsec ($\sim 0.35 $\,pixels), which corresponds to $2.5-3$\,pc 
at the distance of NGC\,3610, i.e. similar to the mean $R_{\rm eff}$ obtained 
with ACS data for GCs in nearer early-type galaxies, like the ones in Virgo and Fornax 
clusters \citep[e.g.][]{jor05,mas10}.

In order to explore relations between size, colour, and luminosity of GCs, the 
left-hand panel of Fig.\,\ref{mvreff} shows $R_{\rm eff}$ versus $M_V$ 
for the 13 clusters in NGC\,3610 spectroscopically confirmed by \citet{str03,str04b}  
(highlighted with open circles), together with GCs/UCDs from 
Virgo \citep{bro11}, Fornax \citep{mie04}, Antlia \citep{cas13a,cas14},
Hydra \citep{mis11}, and Coma \citep{chib11}. The colour bar 
represents $(V-I)_0$ colours. For Virgo objects, the colour was
obtained from $g'i'$-band photometry applying the transformations derived in
this paper. As expected, there seems to be no correlation between mean size and 
luminosity for the $M_V$ range of typical GCs \citep [e.g.][]{puz14}, excluding 
the so called `extended clusters' (EC). The ECs have similar brightness than GCs 
but present larger sizes, i.e. $R_{\rm eff} > 10$\,pc \citep[][and references therein]{bru12,forb13}.
For brighter objects, in the UCD's domain, the size globally increases with 
luminosity \citep{bru12,nor14}.     

Among the NGC\,3610 confirmed clusters \citep{str04b}, and according to the $(V-I)$ colour 
limits adopted in this paper, 2 clusters are blue, 8 clusters  
are red, and 3 have intermediate colours. Table~\ref{rad} gives 
their magnitudes and colours from the $g'i'$-band photometry calculated in this paper, as well as 
their estimated effective radii and corresponding errors. All of them have radii in the range of 
typical GCs with the exception of a single object, spectroscopically confirmed 
and identified as W30 \citep{str03}, that is marginally larger ($R_{\rm eff} = 11.55$\,pc) 
and might  be classified as EC. 
The two young and metal-rich clusters are in the reddest subsample, with sizes 
$R_{\rm eff} = 3.22$ and $7.36$\,pc for W6 and W11, respectively. 

\begin{table}
\begin{center}
\caption{Magnitudes, colours (from GMOS data), and effective radii for the spectroscopically 
confirmed clusters \citep{str04b}. Identifications (ID) are taken from Strader et al. } 
\label{rad}
\begin{tabular}{@{}cccc@{}}   
\hline   
\multicolumn{1}{@{}c}{ID}&\multicolumn{1}{c}{$i'_0$}&\multicolumn{1}{c}{$(g'-i')_0$}&\multicolumn{1}{c}{$R_{\rm eff}$}\\   
\multicolumn{1}{@{}c}{~~}&\multicolumn{1}{c}{(mag)}&\multicolumn{1}{c}{(mag)}&\multicolumn{1}{c}{(pc)}\\   
\hline   
W3  & $20.84\pm0.01$ &  $1.07\pm0.01$ & $2.86\pm0.77$ \\ 
W6  & $21.01\pm0.01$ &  $1.06\pm0.01$ & $3.22\pm0.09$ \\
W9  & $21.09\pm0.02$ &  $0.98\pm0.02$ & $2.97\pm0.09$ \\
W10 & $21.39\pm0.01$ &  $0.92\pm0.02$ & $4.38\pm0.17$ \\
W11 & $21.24\pm0.01$ &  $1.20\pm0.01$ & $7.36\pm0.26$ \\
W12 & $21.48\pm0.01$ &  $0.94\pm0.01$ & $2.37\pm0.26$ \\
W22 & $22.01\pm0.01$ &  $0.89\pm0.02$ & $2.98\pm0.26$ \\
W28 & $22.23\pm0.01$ &  $1.02\pm0.02$ & $3.83\pm0.09$ \\
W30 & $22.39\pm0.01$ &  $1.00\pm0.02$ & $11.55\pm0.34$ \\
W31 & $22.44\pm0.01$ &  $1.15\pm0.02$ & $4.98\pm0.68$ \\
W32 & $22.43\pm0.01$ &  $1.14\pm0.02$ & $3.22\pm0.26$ \\
W33 & $22.60\pm0.01$ &  $0.97\pm0.02$ & $3.47\pm0.60$ \\
W40 & $22.84\pm0.01$ &  $1.16\pm0.02$ & $6.02\pm0.60$ \\
\hline
\end{tabular}  
\end{center}
\end{table} 

The right-hand panel of Fig.\,\ref{mvreff} presents an analogue plot but for NGC\,3610 GC 
candidates, including the 13 spectroscopically confirmed clusters (identified with open circles). 
Several EC candidates are also present, within the luminosity range typical of GCs. 
As quoted above, the cluster named W30 is included among them. This EC has intermediate colours 
and, according to the Lick/IDS analysis performed by \citet{str04b}, it has a typical GC-like age.   
There is a small group of both confirmed objects and candidates on the bright 
side, close to $M_V \sim -11$, which are in the brightness limit between massive GCs 
and UCDs. The young metal-rich clusters W6 and W11 are located in the same place as this 
group and, comparing with the analogous plot presented by \citet{nor14} in their fig.~11, 
they may fall on an extension of the sequence of YMCs towards fainter and smaller clusters.   
As part of this group, there are 3 GC-like objects with available 
$R_{\rm eff}$ measurements and whose positions match with X-ray point-sources 
from \citet{liu11} (identified with open triangles). 
Out of these latter objects, one is the spectroscopically confirmed GC identified as W9 
by \citet{str03}. From the Lick/IDS indices, these authors showed that W9 is old and red  
(metal-rich), with metallicity $[Fe/H] = -1.2\pm0.2$. The two remaining GC candidates 
lack spectroscopic 
confirmation, but their photometric ages and metallicities were derived by \citet{geo12}. 
They correspond to young objects: G8, $\sim 2$\,Gyr and $[Z/H] \sim -0.2$, and G14, 
$\sim 5$\,Gyr and $[Z/H] \sim -0.4$. 

Thus, most of the GC candidates brighter than $I_0 = 24$ have sizes in the expected range for 
GCSs of nearby early-type galaxies, but we also detected a group of ECs. 
Among the 13 spectroscopically confirmed GCs, only one seems to be an EC and the rest have normal 
GC-sizes. In particular, one of the two young and metal-rich ones (W11, $R_{\rm eff} = 7.36$\,pc) is 
marginally larger than the mean. 

\section{Surface photometry of NGC\,3610}

\begin{figure}    
\includegraphics[width=80mm]{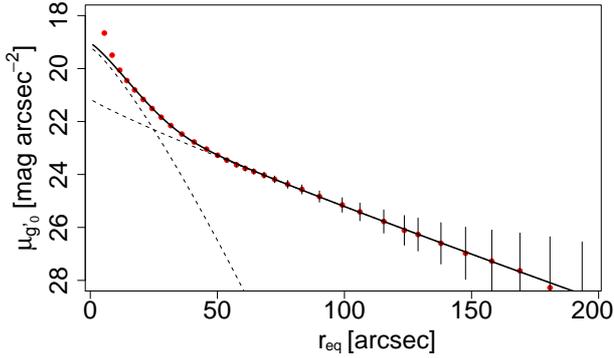}
\caption{Surface-brightness profile of the NGC\,3610, $g'$-band. The 
dashed lines represent two S\'ersic models fitted to the galaxy profile,
while the solid line indicates the sum of them.}    
\label{perfil}    
\end{figure}

Fig.\,\ref{perfil} shows the surface-brightness profile of NGC\,3610 (surface
brightness versus equivalent radius, being $r_{\rm eq} = \sqrt{a~b} = a
\sqrt{1-\epsilon}$, where $a$ is the isophote semi-major axis and
$\epsilon$ its ellipticity) in the $g'$ filter, obtained with the 
\textsc{iraf} task ELLIPSE. The galaxy profile has been fitted with 
two S\'ersic models \citep{ser68} expressed in surface-brightness 
units (mag\,arcsec$^{-2}$)

\begin{equation}  
\mu(r) = \mu_0 + 1.0857\,\left(\frac{r_{\rm eq}}{r_0}\right)^{\frac{1}{n}},  
\end{equation}  
  
\noindent where $\mu_0$ is the central surface-brightness, $r_0$ 
is a scale parameter and and $n$ is the S\'ersic shape index (i.e. 
$n=1$ corresponds to an exponential profile and $n=4$ to a 
de Vaucouleurs profile). The resulting parameters for the inner 
and outer components are listed in Table~\ref{par}, 
where we have also included the respective effective radii,
according to the relation

\begin{equation}  
r_{\rm eff} = b_{\rm n}^{n}r_0 
\end{equation}  

\noindent where $b_{\rm n}$ is a function of the $n$ index, that may be
estimated with the expression given by \citet{cio91}.

\begin{table}
\begin{center}
\caption{Parameters of the two S\'ersic models fitted to the galaxy
profile in the $g'$ filter. Both $r_0$ and $r_{eff}$ are expressed in 
arcsec.}
\label{par}
\begin{tabular}{@{}ccccc@{}}   
\hline   
\multicolumn{1}{@{}c}{Component}&\multicolumn{1}{c}{$\mu_{0}$}&\multicolumn{1}{c}{$r_{0}$}&\multicolumn{1}{c}{$n$}&\multicolumn{1}{c}{$r_{\rm eff}$}\\   
\hline   
Inner & $19.2\pm0.2$ & $10.6\pm1.3$ & $0.8\pm0.08$ & $13.6$\\
Outer & $21.1\pm0.2$ & $23.2\pm2.8$ & $1.1\pm0.05$ & $47.3$\\
\hline
\end{tabular}  
\end{center}
\end{table}   

If we compare our two-component fit with the photometric analysis performed 
by Whitmore et al., we are not able to detect 
their small twisted `inner disc' 
within 3\,arcsec \citep{whi02}. We just attempt to fit 
the inner part excluding the very central 20\,arcsec, where the 
profile gets steeper, as we are mostly interested in the large-scale 
brightness distribution. 

\begin{figure}    
\includegraphics[width=70mm]{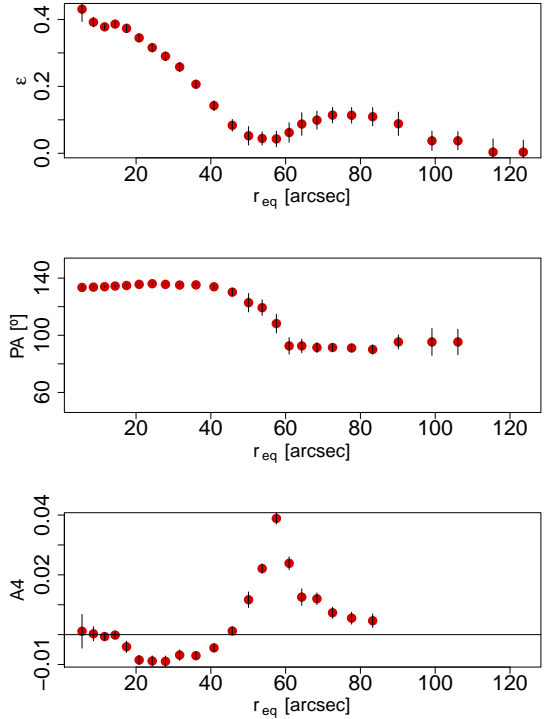}
\caption{Ellipticity $\epsilon$, position angle $PA$, and $A4$ Fourier coefficient 
of the elliptical isophotes versus equivalent radius $r_{\rm eq}$}.    
\label{param}    
\end{figure}

\begin{figure}    
\includegraphics[width=80mm]{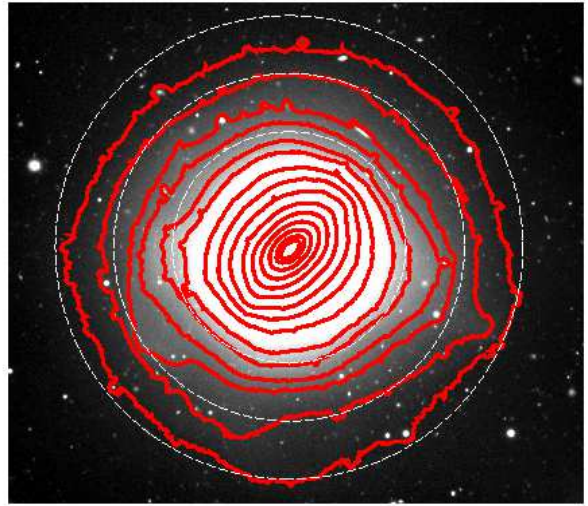}
\caption{GMOS image of NGC\,3610 (N3610F) in the $g'$-band with light contours 
overplotted in red (solid lines). Three concentric circles (dashed lines) 
are shown at 60, 90, and 120\,arcsec from the galaxy centre, in order to 
facilitate the comparison with other figures and the text.}    
\label{isof}    
\end{figure}

Fig.\,\ref{param} shows the parameters obtained with ELLIPSE that 
characterise the fitted elliptical isophotes: ellipticity $\epsilon$, position 
angle $PA$ (measured positive from $N$ to $E$), and the $A4$ Fourier 
coefficient that is related to diskiness ($A4>0$ discy isophotes, $A4<0$ 
boxy isophotes), as a function of $r_{\rm eq}$. 
The surface-brightness contours are presented in Fig.\,\ref{isof} superimposed 
to a $g'$ GMOS image. The behaviour of all these parameters, which can be 
globally followed on Fig.\,\ref{isof}, is a direct consequence of the 
complex structure of the galaxy. The $\epsilon$ decreases from a central 
value of $\sim0.4$ down to $\sim0.05$ while reaching 50\,arcsec. This radial range corresponds 
to the elliptical isophotes of our inner component, i.e. a disc we will describe below.
Further out, $\epsilon$ rises slightly but remaining smaller than $0.1$, and finally 
approaches 0 between 110 and 120\,arcsec. The $PA$ presents a clear variation from $132$ to 
$92$\,deg, that is a change of $\sim40$\,deg between 40 and 60\,arcsec. 
The global behaviour of $\epsilon$ and $PA$ agree with those presented by 
\citet{gou94} within their more limited radius of $\sim 40$\,arcsec.   
Finally, the $A4$ coefficient confirms earlier statements that there are boxy 
isophotes ($A4<0$) in NGC\,3610, as pointed out by \citet{sco90} for instance, which 
is considered as a clear evidence of past mergers. These boxy isophotes are present  
between $\sim$20 (the lower limit of our fit) and $\sim$45\,arcsec and can also be 
identified in Fig.\,\ref{isof}. At larger semi-axes, $A4$ is positive and varies 
continuously reaching a maximum at $\sim60$\,arcsec. Looking at the global  
picture, the behaviour of these parameters agree with the existence   
of a slope change in the light profile between $\sim50$ and $\sim60$\,arcsec, where 
our inner component fades away, and the outer component dominates further out.

According to the characteristics of our inner component (S\'ersic 
index $n\sim 0.8$ and $r_{\rm eff} \sim 13$\,arcsec), it corresponds to the 
inner disc already detected by \citet{sco90} in this galaxy, that we identify 
through the elliptical isophotes for $r_{\rm eq} < 50$\,arcsec, with a $PA = 132$\,deg, 
which agrees perfectly with their $PA$. However, with our GMOS images this inner disc 
can now be traced up to a much larger radius, $\sim 60$\,arcsec. 
This disc is aligned with the small 3\,arcsec twisted `inner disc' detected by \citet{whi97}, 
and they are probably part of the same component whose central region shows a very steep rise 
in the surface-brightness profile, reaching $\mu_{g'0} \sim 19$\,mag\,arcsec$^{-2}$. 
It is interesting to note that the extent we obtain for this disc agrees with the $R_g$ 
where the GC density profile deviates from the power-law (see Fig.\,\ref{drad}) and 
with the radial range where the (inner) GC population is dominated by 
red clusters (see also Fig.\,\ref{espa}), all likely related to the merger event. 

As said above, our GMOS images let us reach further out than previous 
studies and fully characterise the outer component. 
In spite of the low surface-brightness of the outer region, we are able 
to recover a S\'ersic model with a shape index $n \sim1.1$ and a quite 
large $r_{\rm eff}\sim 47$\,arcsec, i.e. a bulge/spheroidal extended component 
associated with the fine-structure present in the surface-brightness distribution.
According to the light profile (Fig.\,\ref{perfil}), it extends up to a 
minimum $R_g \sim 180$\,arcsec where the limiting 
surface-brightness of 28\,mag\,arcsec$^{-2}$ is reached in the $g'$-band. 
It corresponds to the `bulge' modeled by \citet{sco90} up to a much shorter 
$R_g \sim 100$\,arcsec. 
Searching for a correlation between the stellar and GC populations, it is 
worth noting that the GC candidates located between 1 and 2\,arcmin 
(Fig\,\ref{dcol}) present the usual bimodal colour distribution, 
while further out, a small subsample of clusters between 2 and 4\,arcmin 
mainly show up at an `intermediate' colour range (mean $(g'-i')_0 = 0.87$).

\begin{figure*}    
\includegraphics[width=80.0mm]{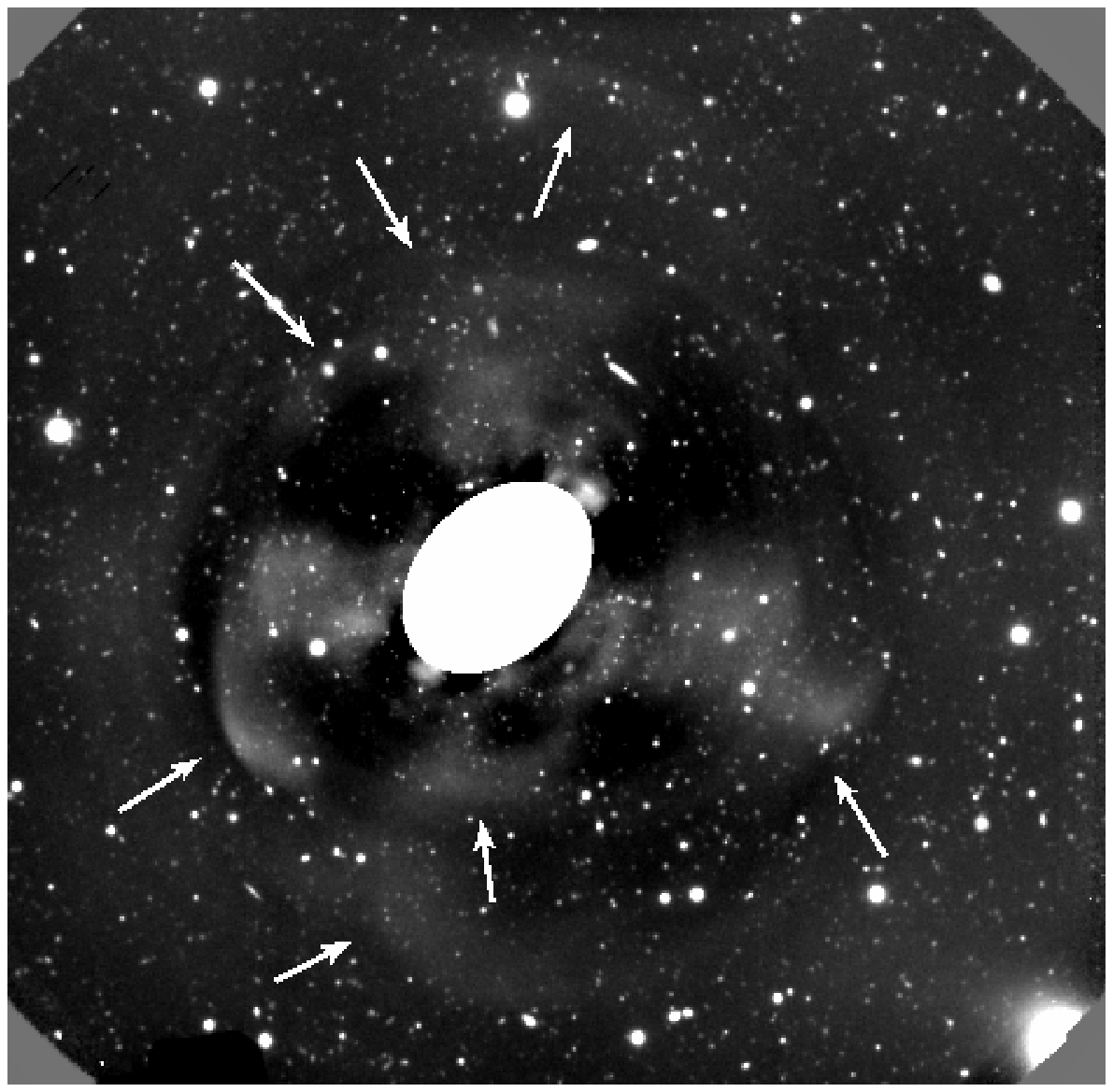}
\includegraphics[width=82.0mm]{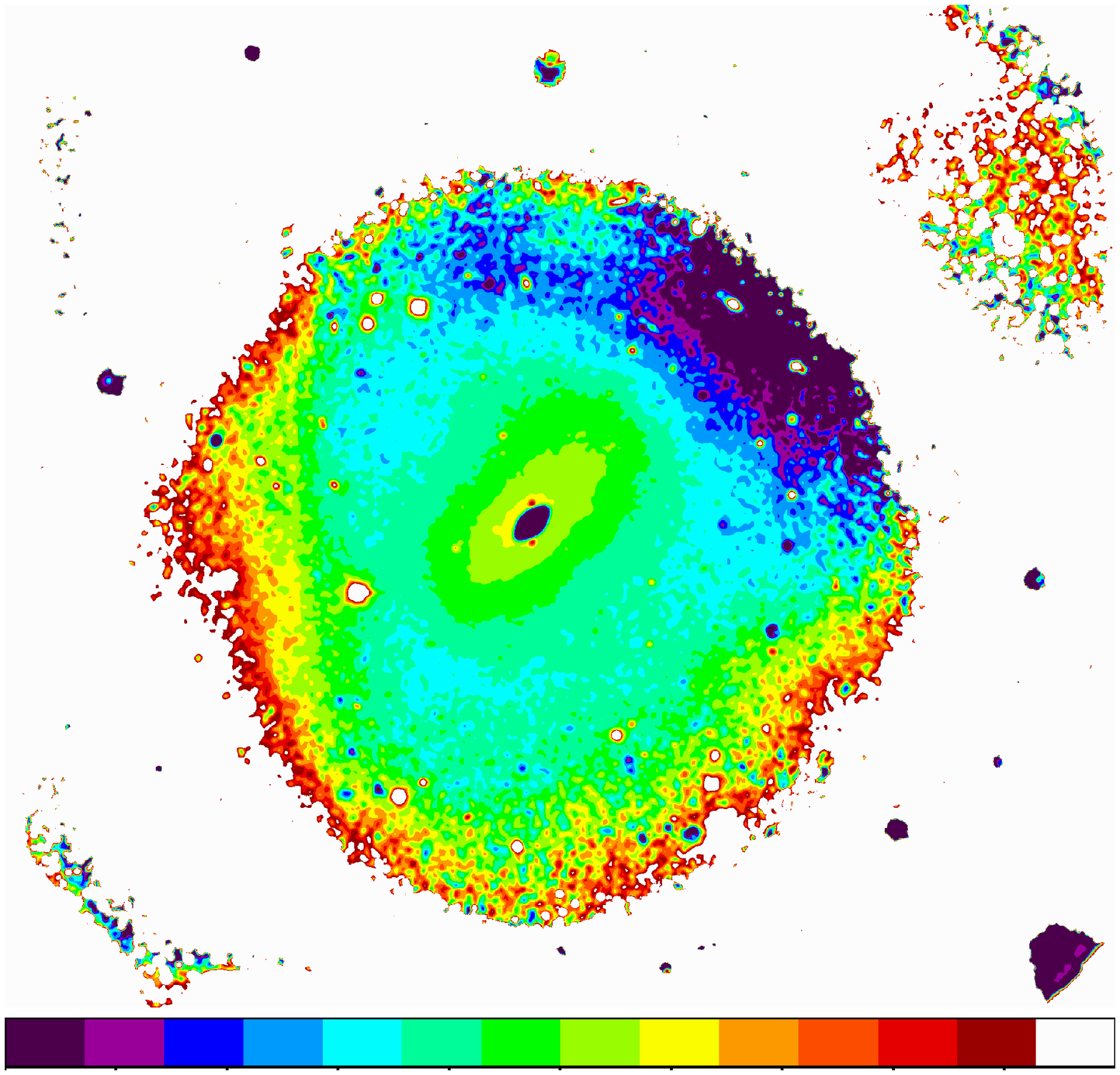}
\caption{Left: GMOS image of NGC\,3610 (N3610F) in the $g'$-band, once the smoothed
component of the galaxy is subtracted. White arrows indicate shells and plumes. 
Right: Smoothed $(g'-i')$ colour map of the NGC\,3610. The colour bar shown at the bottom 
spans from $(g'-i')$ = 0.8 to 1.5}    
\label{resid}    
\end{figure*}

In order to draw a picture of the galaxy's outer structure, we obtained a 
residual map by subtracting from the original $g'$-band image a smoothed 
galaxy model performed with ELLIPSE and BMODEL within \textsc{iraf}, where the 
central region has been masked in order to enhance the external features. 
The resulting image is presented in the left-hand panel of Fig.\,\ref{resid}, 
where a complex underlying structure can be seen with unprecedented detail, 
including a wealth of shells and faint plumes, indicated with arrows, 
and a global X-structure. 
They appear over the whole GMOS image, covering the FOV of 
$5.5\times5.5~{\rm arcmin}^2$, that is, 
taking into account that we are just looking at a projected brightness 
distribution, they seem to be mostly associated to the outer spheroidal 
component quoted above. They are clearly indicating a recent history of 
strong tidal interactions and perturbations. 
 
The $(g'-i')$ colour map depicted on the right-hand panel of Fig.\,\ref{resid} 
shows that there is no axisymmetric distribution, but a large-scale mild 
colour gradient is present in the global stellar population. 
It spans from blue ($(g'-i') \sim$ 0.8) at the $NW$  to red ($(g'-i') 
\sim$ 1.5) towards the opposite $SE$ side, following a similar direction as
the semi-major axes of the inner galaxy (disc) isophotes. No traces of 
dust can be detected in the galaxy body, but the lack of a global radially 
symmetric colour distribution is understood as another piece of evidence of 
a past merger.

\section{Discussion}

It is quite clear that NGC\,3610 is a dynamically young, merger remnant, as shown by the 
plenty of features in the surface-brightness distribution, the inner embedded disc 
(within an elliptical galaxy), the boxy isophotes, the presence of spectroscopically 
confirmed young and metal-rich clusters, etc. 
To this picture we add a new determination of the extent of that inner disc and of the 
spheroidal outer component, much larger than previously stated, a new image of the large-scale 
fine-structure with a detail never achieved before, and a colour map with a visible gradient. 
With respect to the GCs, we also obtained the extent of the whole system in NGC\,3610, i.e. 40\,kpc 
Section\,\ref{projdist}), and analysed them up to this new $ R_g $. The new analysis of the GC 
sub-population for $R_g >$ 1\,arcmin includes a GC bimodal colour distribution between 
1~$< R_g <$ 2\,arcmin and the presence of a small subsample of clusters with `intermediate' colours 
(mean $(g'-i')_0 = 0.87$) hosted by the outer component (2~$< R_g <$ 4\,arcmin). 

In order to gather more information about these outer objects of intermediate colours, that seem to 
be present at $R_g <$ 1\,arcmin too, we can 
compare their mean colour with theoretical models of single stellar populations (SSPs), 
as we have already done for NGC\,1316 and NGC\,4753 \citep{ric12,cas15a}, which presented 
similar intermediate-colour cluster sub-populations. 
Considering the models performed by \citet{bre12}, assuming SSPs with abundances 
$0.5 - 1 Z_{\sun}$, a \citet{cha01} lognormal IMF, and absorption 
$A_V=$ 0.03 (NED), the mean colour of these clusters corresponds to a population of
1.5 -- 3\,Gyr. These values are consistent with the ages and metallicities obtained by 
\citet{str03,str04b} for the spectroscopically confirmed young clusters W6 and W11, 
though they are located much closer to the galaxy, at $R_g = 22.7$ and 13.7\,arcsec, 
respectively \citep{str04b}. At first, we cannot exclude the idea that the outer cluster 
sub-population of NGC\,3610, with intermediate colours, may be older and more metal-poor objects. 
However, all the observational evidence points to NGC\,3610 being an `intermediate-age' merger 
remnant as, for instance, all the fine-structure present in the surface-brightness distribution 
 as well as the colour gradient detectable in the galaxy stellar 
population, would have disappeared if the merger were much older. With this 
idea in mind, we can speculate wether the sub-population of clusters with intermediate colours 
is associate to an intense burst (or series of bursts) of stars and cluster formation, that 
were triggered by a recent major merger of two gas-rich discs. Such major merger has already 
been proposed by, among others, \citet{str03} who also explained why it is not possible to 
consider the infall of a minor galaxy as the origin of NGC\,3610.

A different paradigm has been supported by \citet{sco95}, who studied a sample of `discy ellipticals',   
including NGC\,3610 and proposed that the embedded stellar discs are not likely the result of
an accretion or merger event, but are primordial. However, through their data and models they 
could not prove that this interpretation was unique \citep[e.g.][]{dej04}. Another discy 
elliptical integrating that sample was NGC\,4660, which has a strong disc. \citet{kem16} have 
recently reported that 
NGC\,4660 has a long tidal filament which implies that this galaxy has undergone a tidal 
interaction, probably a major merger, in the last few Gyr. Moreover, Kemp et al. show that the filament 
may correspond to a residual from star formation induced in the disc by the interaction, 1--2\,Gyr ago. 
In a similar way, it is hard to explain the properties of NGC\,3610 and its GCS described above, 
in the context of a primordial embedded disc. On the contrary, the fact that NGC\,3610 belongs to a 
loose galaxy group may have provided the suitable conditions for a major merger to take place. 

Regarding the fine-structure, the presence of shells in galaxies is considered as an evidence of a 
recent merger, as the shells 
are stellar remnants of the pre-merger systems. \citet{sik06} studied the GCSs in a sample of six 
shell ellipticals, looking for relations between the merger and formation history of the GCs. 
They found evidence of a GC population associated with a recent merger in two of them. 
Afterwards, \citet{sik07} focused on study the shell structures in these galaxies. They concluded 
that the shell distributions are better described by merger models, in detriment of tidal 
interactions \citep[e.g.][]{tho90} or asymmetric local star formation \citep[e.g.][]{loe87}.

\section{Summary and concluding remarks}    

We investigated the dynamically young merger remnant NGC\,3610 and its GCS. 
This massive galaxy is located in a low-density environment and 
has a quite rich fine-structure. Thanks to the large area covered by our GMOS 
data, complemented with archival ACS data from the innermost regions, we are able 
to study both the GCS and the galaxy, in their whole extent.
Our main results are as follows:

\begin{itemize}    

\item[-] The GC colour distribution obtained from GMOS data can be separated in three radial 
ranges: the inner one (30\,arcsec $ < R_g < $ 1\,arcmin) is dominated by red GCs in agreement 
with the results obtained from ACS data; the intermediate range (1~$ < R_g <$ 2\,arcmin) presents 
the typical 
bimodal GC distribution, i.e. old metal-poor and metal-rich GCs that, for instance, are expected 
to have inhabited the pre-merger discs; and the outer region (2~$< R_g <$ 4\,arcmin) has a small 
sub-population 
of clusters, most of them have `intermediate' colours, i.e. between those of the blue and red GCs  
detected in the intermediate radial range. 
According to SSP models, this outer sub-population may correspond to younger clusters originated 
in the merger, 1 -- 3\,Gyr ago, and a bunch of them may be present at smaller radii too. 

\item[-] The extent of the whole GCS, assuming this limit as the galactocentric 
distance at which the background-corrected surface density is equal to the 30 per cent of 
the background level, is calculated as 40\,kpc. The red and blue GC radial profiles are fitted 
by power-laws, being the red profile steeper than the blue one. The inner region, where 
the profile gets flatter and departs from a power-law, is better fitted with a Hubble model. 
The effect of GC erosion is a likely explanation to this latter behaviour.  

\item[-] By means of the GCLF, we calculate a poor old GC population of about $500\pm110$\,members,
that corresponds to a low specific frequency $S_N \sim$\,0.8. Such low $S_N$ is more frequent 
for spiral galaxies 
of similar brightness as NGC\,3610 than for ellipticals, what can be understood as another hint 
pointing to a recent disc-disc merger. 

\item[-] We determine the effective radii of a sample of GC candidates located in the ACS field and  
 the spectroscopically confirmed clusters from \citet{str04b}. According to their sizes 
and/or brightness, the two confirmed young metal-rich clusters are in the limit between GCs and UCDs.

\item[-] From the analysis of the galaxy surface-brightness profile, we confirm the presence of 
an inner embedded disc and clear boxy isophotes. We perform new determinations of the extent 
of such inner disc (10\,kpc) and the outer spheroidal component (30\,kpc). We also obtain a 
residual map that shows the fine-structure with a degree of detail never achieved before.  
Finally, a colour gradient is detected on the galaxy colour map. 

All our results support the interpretation of NGC\,3610 as a recent merger remnant, probably a 
major disc-disc one, with a cluster population integrated by {\it bona-fide} old GCs presumably born in the 
discs, as well as younger clusters originated in the merger, that can be detected at different 
radii. Moreover, we can speculate whether a two-pass merger may have occurred, with more gaseous
dissipation on the first pass that deposited slightly older (redder) clusters nearer the centre, 
and a second pass that originated the clusters with `intermediate' colours in the outer region. 
We expect to study spectroscopically these few `intermediate' colour clusters found 
in the outskirts, in the near future. 

\end{itemize}

\section*{Acknowledgments}
We appreciated the insightful and constructive comments of the referee, which improved this paper.
This work was funded with grants from Consejo Nacional de Investigaciones   
Cient\'{\i}ficas y T\'ecnicas de la Rep\'ublica Argentina (PIP 112-201101-00393), Agencia Nacional de   
Promoci\'on Cient\'{\i}fica y Tecnol\'ogica (PICT-2013-0317), and Universidad Nacional de La Plata 
(UNLP 11-G124), Argentina. \\
Based on observations obtained at 
the Gemini Observatory (GN2013A-Q-42), which is operated by the Association of Universities 
for Research in Astronomy, Inc., under a cooperative agreement with the NSF on 
behalf of the Gemini partnership: the National Science Foundation (United 
States), the National Research Council (Canada), CONICYT (Chile), the 
Australian Research Council (Australia), Minist\'{e}rio da Ci\^{e}ncia, 
Tecnologia e Inova\c{c}\~{a}o (Brazil) and Ministerio de Ciencia, 
Tecnolog\'{i}a e Innovaci\'{o}n Productiva (Argentina). Also based on observations 
made with the NASA/ESA Hubble Space Telescope, obtained from the data archive at 
the Space Telescope Science Institute. STScI is operated by the Association of 
Universities for Research in Astronomy, Inc. under NASA contract NAS 5-26555.
This research has made use of the NASA/IPAC Extragalactic Database (NED) which 
is operated by the Jet Propulsion Laboratory, California Institute of Technology, 
under contract with the National Aeronautics and Space Administration.

\bibliographystyle{mnras}
\bibliography{biblio}

% Don't change these lines
\bsp	% typesetting comment
\label{lastpage}
\end{document}